\documentclass[reprint,superscriptaddress,amssymb,aps, pra, twocolumn,graphicx]{revtex4-1}
\usepackage{lipsum}
\usepackage{subfigure}
\usepackage{amsfonts}
\usepackage{amsmath}
\usepackage{txfonts}
\usepackage{amssymb}
\usepackage{amsbsy} 
\usepackage{epsfig}
\usepackage{graphicx}
\usepackage{epstopdf}
\usepackage{geometry}
\usepackage{hyperref}
\usepackage{setspace}

\begin{document}

\title{Simultaneous ground-state cooling of multiple degenerate mechanical modes through cross-Kerr effect }%

\author{Pengyu Wen}
\email{These authors have equal contribution to this work.}
\affiliation{Department of Physics and State Key Laboratory of Low-Dimensional Quantum Physics, Tsinghua University, Beijing 100084, China}

\author{Xuan Mao}
\email{These authors have equal contribution to this work.}
\affiliation{Department of Physics and State Key Laboratory of Low-Dimensional Quantum Physics, Tsinghua University, Beijing 100084, China}

\author{Min Wang}
\affiliation{Beijing Academy of Quantum Information Sciences, Beijing 100193, China}

\author{Chuan Wang}
\affiliation{School of Artificial Intelligence, Beijing Normal University, Beijing 100875, China}

\author{Gui-Qin Li}
\affiliation{Department of Physics and State Key Laboratory of Low-Dimensional Quantum Physics, Tsinghua University, Beijing 100084, China}
\affiliation{Frontier Science Center for Quantum Information, Beijing 100084, China}

\author{Gui-Lu Long}
\email{gllong@tsinghua.edu.cn}
\affiliation{Department of Physics and State Key Laboratory of Low-Dimensional Quantum Physics, Tsinghua University, Beijing 100084, China}
\affiliation{Beijing Academy of Quantum Information Sciences, Beijing 100193, China}
\affiliation{Frontier Science Center for Quantum Information, Beijing 100084, China}

\date{\today}%

\begin{abstract}
  Simultaneous ground-state cooling of multiple degenerate mechanical modes is a tough issue in optomechanical system due to the existence of the dark mode effect. Here we propose a universal and scalable method to break the dark mode effect of two degenerate mechanical modes by introducing the cross-Kerr (CK) nonlinearity. At most four stable steady states can be achieved in our scheme in the presence of the CK effect, different from the bistable behavior of the standard optomechanical system. Under the constant input laser power, the effective detuning and mechanical resonant frequency can be modulated by the CK nonlinearity, which results in an optimal CK coupling strength for cooling. Similarly, there will be an optimal input laser power for cooling when the CK coupling strength stays fixed. Our scheme can be extended to break the dark mode effect of multiple degenerate mechanical modes by introducing more than one CK effects. To fulfill the requirement of the simultaneous ground-state cooling of N multiple degenerate mechanical modes N-1 CK effects with different strengths are needed. Our proposal provides new insights in dark mode control and might pave the way to manipulating of multiple quantum states in macroscopic system.

\end{abstract}
\maketitle

\section{Introduction}
      Optomechanics\cite{aspelmeyer2014cavity}, studying the interaction between the light and mechanical vibrations, has been proved as a fertile area for quantum information processing\cite{stannigel2012optomechanical, stannigel2011optomechanical,barzanjeh2022optomechanics, dong2021unconventional}, high-performance sensing\cite{chen2017exceptional, zhi2017single,lai2019observation, massel2011microwave,huang2013demonstration,qin2021experimental,wan2018experimental, mao2022experimental, mao2020enhanced, jing2018nanoparticle,yin2011three,yang2022high} and investigating macroscopic quantum phenomena\cite{ockeloen2018stabilized, riedinger2018remote, ockeloen2016quantum,liao2016macroscopic}, etc. To make these applications realizable a pivotal issue is to suppress the thermal noise of phonons, i.e. the ground-state cooling of mechanical modes. Although the ground-state cooling of single mechanical mode has been proposed in theory\cite{wilson2007theory,genes2008ground,liu2013dynamic,marquardt2007quantum,xia2009ground,tian2009ground,wang2019breaking} and demonstrated in experiments\cite{park2009resolved,schliesser2008resolved, groblacher2008radiation, riviere2011optomechanical,xu2019nonreciprocal}, the simultaneous cooling of multiple degenerate mechanical modes is still a stubborn obstacle because of the forming of dark mode\cite{massel2012multimode,sommer2019partial,shkarin2014optically} . Theoretical proposals for breaking the dark-mode effect has been made by introducing nonreciprocal coupling\cite{lai2020nonreciprocal}, auxiliary-cavity-mode\cite{huang2021multimode} and multiple optical modes\cite{liu2022ground}.

      Nonlinearity has played an important role in the generation of quantum superposition of states \cite{marshall2003towards, vitali2007optomechanical, lu2018entanglement} and quantum squeezed states\cite{clerk2008back, lu2015steady}. Different from the well-known Kerr effect which resulting from the change of the refractive index of a nonlinear medium, the cross-Kerr (CK) effect can be realized in the superconducting circuit \cite{khan2015cross,heikkila2014enhancing} or the atomic system with electromagnetically induced transparency \cite{yang2009enhanced,sinclair2019observation}. It has been shown that the CK effect facilitates the creation of entanglement between photons\cite{bartkowiak2014quantum, sheng2012efficient}, attaining the strong coupling\cite{johansson2014optomechanical, ludwig2012enhanced, xiong2016cross} and quantum information processing\cite{sheng2008efficient,dong2008generation}.

      Inspired by the sideband shift induced by the CK effect\cite{khan2015cross}, we decide to utilize this nonlinearity to break the dark-mode effect. We consider an optomechanical system consisting of two degenerate mechanical modes coupling to one optical mode. We find that there are at most four stable steady states, which is more complex than the bistable behavior of standard optomechanical system since we have extra CK nonlinearity here. By using the steady-state result of the first branch, the simultaneous ground-state cooling of two mechanical modes can be achieved when there is CK coupling between one mechanical mode and the cavity mode. Moreover, there will be an optimal input laser power (CK coupling strength) for cooling when CK coupling strength (input laser power) is fixed due to the modification of the effective detuing and the mechanical resonant frequency. We also extend our scheme to system with more than two degenerate mechanical modes and find that the simultaneous cooling of N degenerate mechanical modes is accessible only when the cavity mode has CK coupling with N-1 mechanical modes whose strengths are different. Our scheme  shows its potential for the engineering of dark-mode effect and  quantum state in macroscopic systems.

      The paper is organized as follows. In section \ref{section2} we describe the model and solve the dynamical equation. In section \ref{section31} we analyze the steady-state behavior. In section \ref{section32} we discuss the cooling effect and the extension of the scheme. The conclusion is given in section \ref{con}. 

\section{Model and Method \label{section2}}

\begin{figure}
  \includegraphics[width=0.8\linewidth]{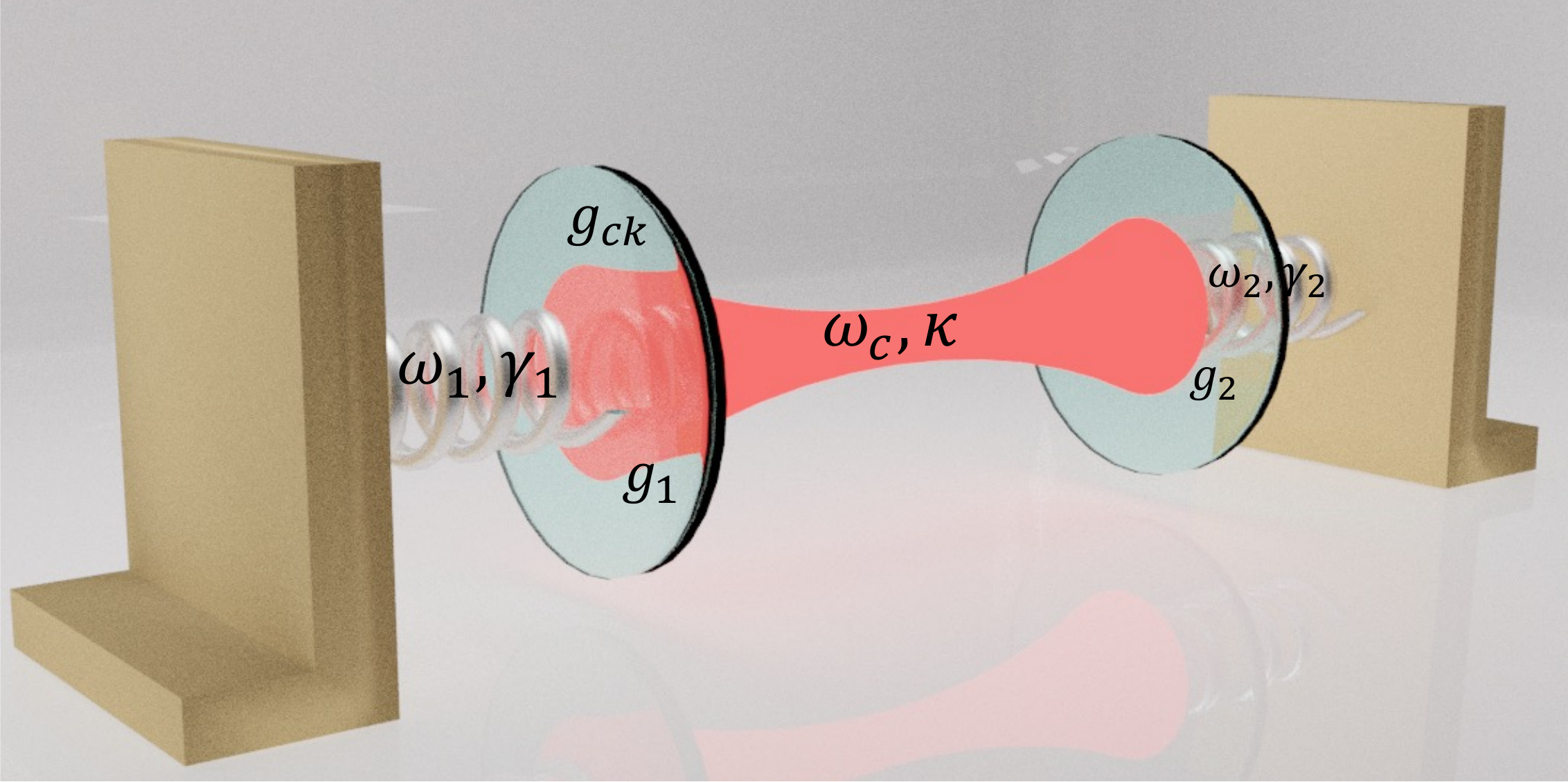}
  \caption{Scheme of optomechanical system with CK effect. The cavity mode is coupled to two degenerate mechanical modes through standard optomechanical interaction. Additionally, there exists CK effect between mechanical mode 1 and the cavity mode. }
  \label{model}
  \end{figure}
  We consider an optomechanical system with CK effect as shown in Fig.\ref{model}. Two degenerate mechanical modes, i.e. the frequency and the linewidth of whom satisfy $\omega_{1}=\omega_{2}=\omega_{m},\gamma_{1}=\gamma_{2}$, are coupled to the cavity mode via standard optomechanical interaction. Specifically, we allow the CK coupling between the cavity mode and the mechanical mode 1. The Hamiltonian in the rotation frame with frequency $\omega_{L}$ of the system is 
  \begin{align}
    H=& \Delta a^{\dagger}a+ \omega_{1}b^{\dagger}_{1}b_{1}+ \omega_{2}b^{\dagger}_{2}b_{2}-g_{1}a^{\dagger}a(b^{\dagger}_{1}+b_{1})\nonumber \\ &-g_{2}a^{\dagger}a(b^{\dagger}_{2}+b_{2}) -g_{ck} a^{\dagger}a b^{\dagger}_{1}b_{1}+i (\epsilon a^{\dagger}-\epsilon^{*}a).
    \end{align}
    
  $a^{\dagger}(a)$,$b^{\dagger}_{1}(b_{1})$ and $b^{\dagger}_{2}(b_{2})$ are the creation (annihilation) operator of the cavity mode, mechanical mode 1 and mechanical mode 2, respectively.    $\Delta=\omega_{c}-\omega_{L}$ represents the detuning between the cavity frequency and the driving frequency. $g_{1}(g_{2})$ denotes single photon coupling strength between the cavity mode and the mechanical mode 1 (2). $g_{ck}$ is the CK coupling strength. $\epsilon=\sqrt{2 \kappa P/\hbar \omega_{L}}$ is the driving laser amplitude with $\kappa$ representing the decay of the cavity mode  and $P$ denoting the input driving laser power. 
  
  Using the Heisenberg-Langevin approach, the dynamical quantum Langevin equations are
  \begin{align}
    \dot{a}=&(-i\Delta-\kappa)a+i g_{1} a(b^{\dagger}_{1}+b_{1})+ i g_{2} a(b^{\dagger}_{2}+b_{2}) +i g_{ck} a b^{\dagger}_{1}b_{1}\nonumber \\  &+\epsilon+\sqrt{2\kappa} a_{in}, \nonumber \\
    \dot{b}_{1}=&(-i\omega_{1}-\gamma_{1})b_{1}+ig_{1}a^{\dagger}a+ig_{ck}a^{\dagger}a b_{1} +\sqrt{2\gamma_{1}}b_{1,in}, \\
    \dot{b}_{2}=&(-i\omega_{2}-\gamma_{2})b_{2}+ig_{2}a^{\dagger}a +\sqrt{2\gamma_{2}}b_{2,in}.\nonumber
    \end{align}

   $a_{in}$ and $b_{1,in}(b_{2,in})$ are the input noise channel of the cavity mode and mechanical mode 1 (2), respectively. Utilizing the Markov approximation, the time correlation functions of these noise operators are  
   \begin{align}
    <a_{in}(t)a^{\dagger}_{in}(t^{\prime})>&=\delta(t-t^{\prime}),\nonumber \\ 
    <b^{\dagger}_{in}(t)b_{in}(t^{\prime})>&=n_{th}\delta(t-t^{\prime}), \\
    <b_{in}(t)b^{\dagger}_{in}(t^{\prime})>&=(n_{th}+1)\delta(t-t^{\prime}),\nonumber
    \end{align}
  where $n_{th1(2)}=(e^{\hbar \omega_{1(2)}/k_{B}T}-1)^{-1}$ with $T$ denoting the bath temperature. Following the traditional linearized approach $a=\alpha+\delta a$,$b_{1(2)}=\beta_{1(2)}+\delta b_{1(2)}$ we have the steady-state solution as 
  \begin{align}
    &\big[i(\Delta-2g_{1}Re(\beta_{1})-2g_{2}Re(\beta_{2})-g_{ck}\vert \beta_{1}\vert^{2} )+\kappa\big]\alpha=\epsilon,\nonumber \\ 
    &\big[i(\omega_{1}-g_{ck}\vert \alpha\vert^{2})+\gamma_{1}\big]\beta_{1}=ig_{1}\vert \alpha\vert^{2},\label{steadyeq} \\
    &(i\omega_{2}+\gamma_{2})\beta_{2}=ig_{2}\vert \alpha\vert^{2},\nonumber
    \end{align}
    \begin{figure}
      \includegraphics[width=\linewidth]{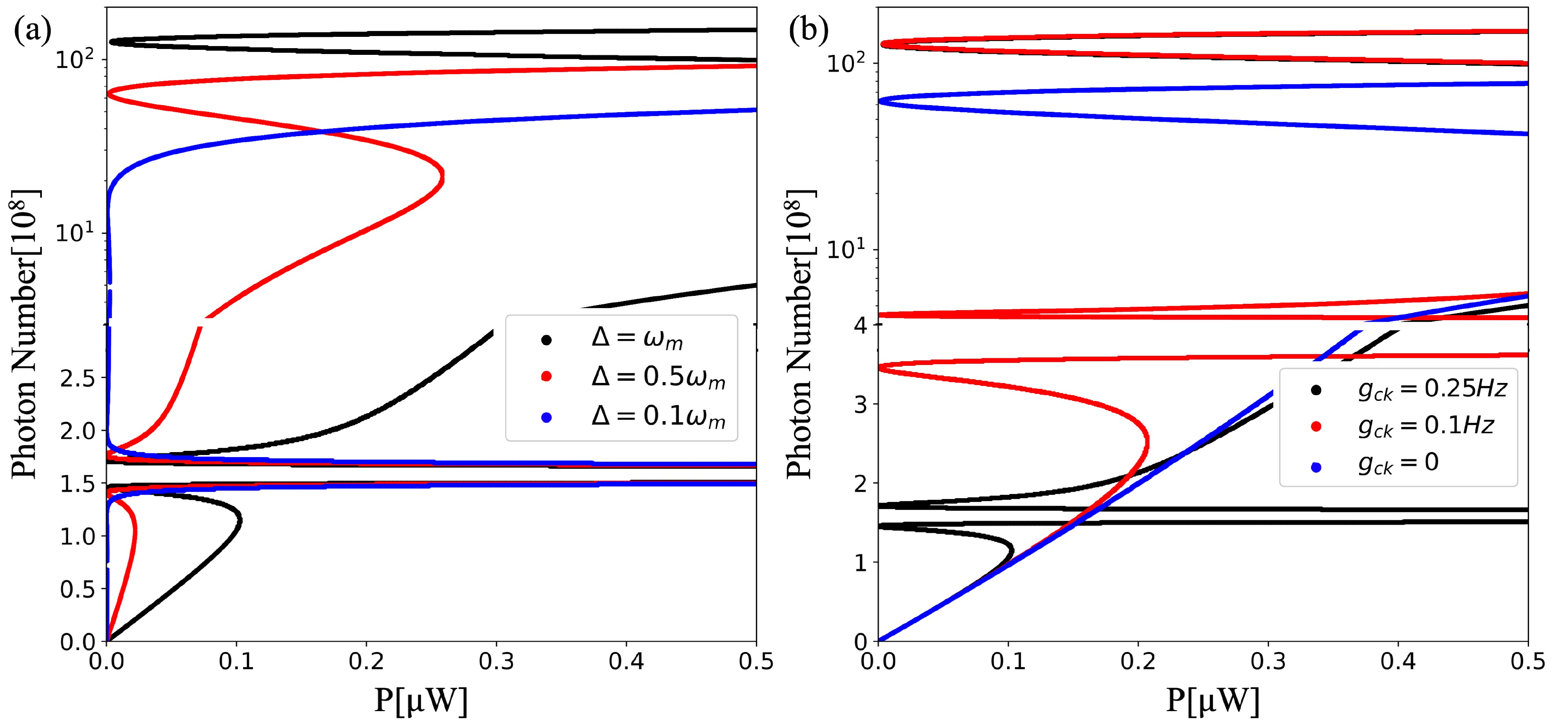}
      \caption{(a) Mean photon number $\vert\alpha\vert^{2}$ with different detuning varies as input driving laser power.$g_{ck}=0.25$Hz (b) Mean photon number $\vert\alpha\vert^{2}$ with different  $g_{ck}$ varies as input driving laser power. $\Delta=\omega_{m}$ Other parameters are $\omega_{1}=\omega_{2}=\omega_{m}=2\pi\times6.3$MHz,$\gamma_{1}=\gamma_{2}=40$Hz,$g_{1}=g_{2}=250$Hz,$\kappa=2\pi\times0.1$MHz,$\omega_{L}=2\pi\times1.3$GHz. Note that the vertical axises of both (a) and (b) are split into two parts, where the upper vertical axis is logarithmic. }
      \label{multistates}
      \end{figure}
    and fluctuation equations as 
    \begin{align}
      \delta \dot{a}=&(-i\Delta^{\prime} -\kappa)\delta a +i g_{1}\alpha(\delta b_{1}+\delta b^{\dagger}_{1})+i g_{2}\alpha(\delta b_{2}+\delta b^{\dagger}_{2})\nonumber\\ & +i g_{ck}(\alpha \beta_{1} \delta b^{\dagger}_{1}+\alpha \beta^{*}_{1} \delta b_{1})+\sqrt{2\kappa}a_{in},\nonumber \\ 
      \delta \dot{b}_{1}=&(-i \Omega_{1}-\Gamma_{1})\delta b_{1}+i g_{1}(\alpha \delta a^{\dagger}+\alpha^{*}\delta a)\nonumber\\ &+i g_{ck}(\alpha \beta_{1}\delta a^{\dagger}+\alpha^{*} \beta_{1}\delta a)+\sqrt{2\gamma_{1}}b_{1,in}, \label{fluceq} \\
      \delta \dot{b}_{2}=&(-i \omega_{2}-\Gamma_{2})\delta b_{2}+i g_{2}(\alpha \delta a^{\dagger}+\alpha^{*}\delta a)+\sqrt{2\gamma_{2}}b_{2,in},\nonumber
      \end{align}
      
      where 
      \begin{align}
        \Delta^{\prime}&=\Delta-2g_{1}Re(\beta_{1})-2g_{2}Re(\beta_{2})-g_{ck}\vert\beta_{1}\vert^{2}\nonumber\\
        \Omega_{1}&=\omega_{1}-g_{ck}\vert\alpha\vert^{2}\label{deltaprime}
        \end{align}
        
        $\Delta^{\prime}$ represents the effective detuning and $\Omega_{1}$ denotes the effective frequency of mechanical mode 1.

\section{Ground-state cooling of the system}
\subsection{Steady-state behavior\label{section31}}

  By analyzing Eq.\ref{steadyeq} we can express $\beta_{1}$ and $\beta_{2}$ as

  \begin{align}
    \beta_{1}&=\frac{ig_{1}\vert\alpha\vert^{2}}{i\Omega_{1}+\gamma_{1}},\nonumber \\ 
    \beta_{2}&=\frac{ig_{2}\vert\alpha\vert^{2}}{i\omega_{2}+\gamma_{2}}. 
    \end{align}

  Substituting  $\beta_{1}$ and $\beta_{2}$ into the first equation of Eq.\ref{steadyeq} we have the steady photon number $\vert\alpha\vert^{2}$ as 

\begin{align}
  \vert\alpha\vert^{2}\bigg[\kappa^{2}+\bigg(\Delta-\frac{2 g^{2}_{1} \vert\alpha\vert^{2} \Omega_{1}}{\Omega^{2}_{1}+\gamma^{2}_{1}}-\frac{2 g^{2}_{2} \vert\alpha\vert^{2} \omega_{2}}{\omega^{2}_{2}+\gamma^{2}_{2}}-\frac{g_{ck}g^{2}_{1}\vert\alpha\vert^{4}}{\Omega^{2}_{1}+\gamma^{2}_{1}}\bigg)^{2}\bigg]=\epsilon^{2}. \label{steadyalpha}
    \end{align}
    In the absence of  $g_{ck}$, Eq.\ref{steadyalpha} becomes a cubic equation which leads to well-known bistable behavior of optomechanical system\cite{ghobadi2011quantum}, which is demonstrated by blue curve in Fig.\ref{multistates}(b). One can see that this blue curve is split into 3 branches. From bottom to top, the first and the third branch denote the stable steady states while the second branch represents the unstable steady-state. When $g_{ck}\neq 0$, i.e. the CK effect exists, one can find that Eq.\ref{steadyalpha} is a seventh-order equation of $\vert\alpha\vert^{2}$ which might lead to at most four stable steady states as shown in other curves in Fig.\ref{multistates} (a) and (b).

  \subsection{Ground-state cooling by breaking dark mode\label{section32}}
  One can simplify Eq.\ref{fluceq} as the form of 
  \begin{align}
    \dot{\mathbf{u}}(t)=\mathbf{Au}(t)+\mathbf{N}(t),
    \end{align}
    where
    \begin{widetext}
    \begin{align}
      \mathbf{u}(t)&=\big[\delta a(t),\delta b_{1}(t),\delta b_{2}(t),\delta a^{\dagger}(t),\delta b^{\dagger}_{1}(t),\delta b^{\dagger}_{2}(t)\big]^{T} ,\\
      \mathbf{N}(t)&=\bigg[\sqrt{2\kappa}  a_{in}(t),\sqrt{2\gamma_{1}} b_{1,in}(t),\sqrt{2\gamma_{2}} b_{2,in}(t),\sqrt{2\kappa} a^{\dagger}_{in}(t),\sqrt{2\gamma_{1}} b^{\dagger}_{1,in}(t),\sqrt{2\gamma_{2}} b^{\dagger}_{2,in}(t)\bigg]^{T},
      \end{align}
    
    and
    
    \begin{equation}
     \mathbf{A}=\\  
        \left(
        \begin{array}{cccccc}
             -i \Delta^{\prime}-\kappa&i g_{1}\alpha+i g_{ck} \alpha \beta^{*}_{1} &i g_{2}\alpha &0&i g_{1}\alpha+i g_{ck} \alpha \beta_{1}&i g_{2}\alpha \\
             i g_{1}\alpha^{*}+i g_{ck} \alpha^{*} \beta_{1}&-i \Omega_{1}-\gamma_{1}&0 &i g_{1}\alpha+i g_{ck} \alpha \beta_{1}&0&0\\
             i g_{2}\alpha^{*}&0&-i \omega_{2}-\gamma_{2} &i g_{2}\alpha&0&0\\
             0&-i g_{1}\alpha^{*}-i g_{ck} \alpha^{*} \beta^{*}_{1}&-i g_{2}\alpha^{*} &i \Delta^{\prime}-\kappa&-i g_{1}\alpha^{*}-i g_{ck} \alpha^{*} \beta_{1}&-i g_{2}\alpha^{*}\\
             -i g_{1}\alpha^{*}-i g_{ck} \alpha^{*} \beta^{*}_{1}&0&0 &-i g_{1}\alpha-i g_{ck} \alpha \beta^{*}_{1}&i \Omega_{1}-\gamma_{1}&0\\
             -i g_{2}\alpha^{*}&0&0 &-i g_{2}\alpha&0&i \omega_{2}-\gamma_{2}\\
        \end{array}
        \right),
        \end{equation}
      \end{widetext}
    with $\mathbf{u}(t)$ denoting fluctuation operator vector,  $\mathbf{N}(t)$ denoting noise operator vector and $\mathbf{A}$ representing coefficient matrix. To get the final steady-state phonon number for each mechanical mode, we only have to solve the Lyapunov equation
    \begin{align}
      \mathbf{AV}+\mathbf{V}\mathbf{A}^{T}=-\mathbf{Q}, \label{lyapunoveq}
      \end{align}
      where $\mathbf{V}$ denoting the covariance matrix with its elements as 
      \begin{align}
        \mathbf{V}_{ij}=\frac{1}{2}[\langle  \mathbf{u}_{i}(\infty)\mathbf{u}_{j}(\infty)\rangle +\langle  \mathbf{u}_{j}(\infty)\mathbf{u}_{i}(\infty)\rangle ],         
        \end{align}
        with $i,j=1-6$. $\mathbf{Q}$ is defined as $\mathbf{Q}=\frac{1}{2}(\mathbf{C}+\mathbf{C}^{T})$ where noise correlation matrix $\mathbf{C}$ is defined as 
        \begin{align}
          \mathbf{C}_{k,l}\delta(t-t^{\prime})=\langle \mathbf{N}_{k}(t)\mathbf{N}_{l}(t^{\prime}) \rangle .        
          \end{align}
  Based on Eq.\ref{lyapunoveq} the final phonon number for each mechanical mode  is obtained as 
  \begin{align}
    n_{1f}&=\langle \delta b^{\dagger}_{1}\delta b_{1} \rangle=\mathbf{V}_{52}-\frac{1}{2},\\
    n_{2f}&=\langle \delta b^{\dagger}_{2}\delta b_{2} \rangle=\mathbf{V}_{63}-\frac{1}{2}.       
    \end{align}
    Note that the solution of Eq.\ref{lyapunoveq} is reliable only when the parameters in $\mathbf{A}$ satisfy the Routh-Hurwitz criterion\cite{grandshteyn1980table}, namely the real parts of all the eigenvalues of $\mathbf{A}$ are negative. The steady-state result, i.e. $\vert \alpha \vert^{2}$  is adopted from the first branch of those curves in Fig.\ref{multistates}.
  \begin{figure}
    \includegraphics[width=\linewidth]{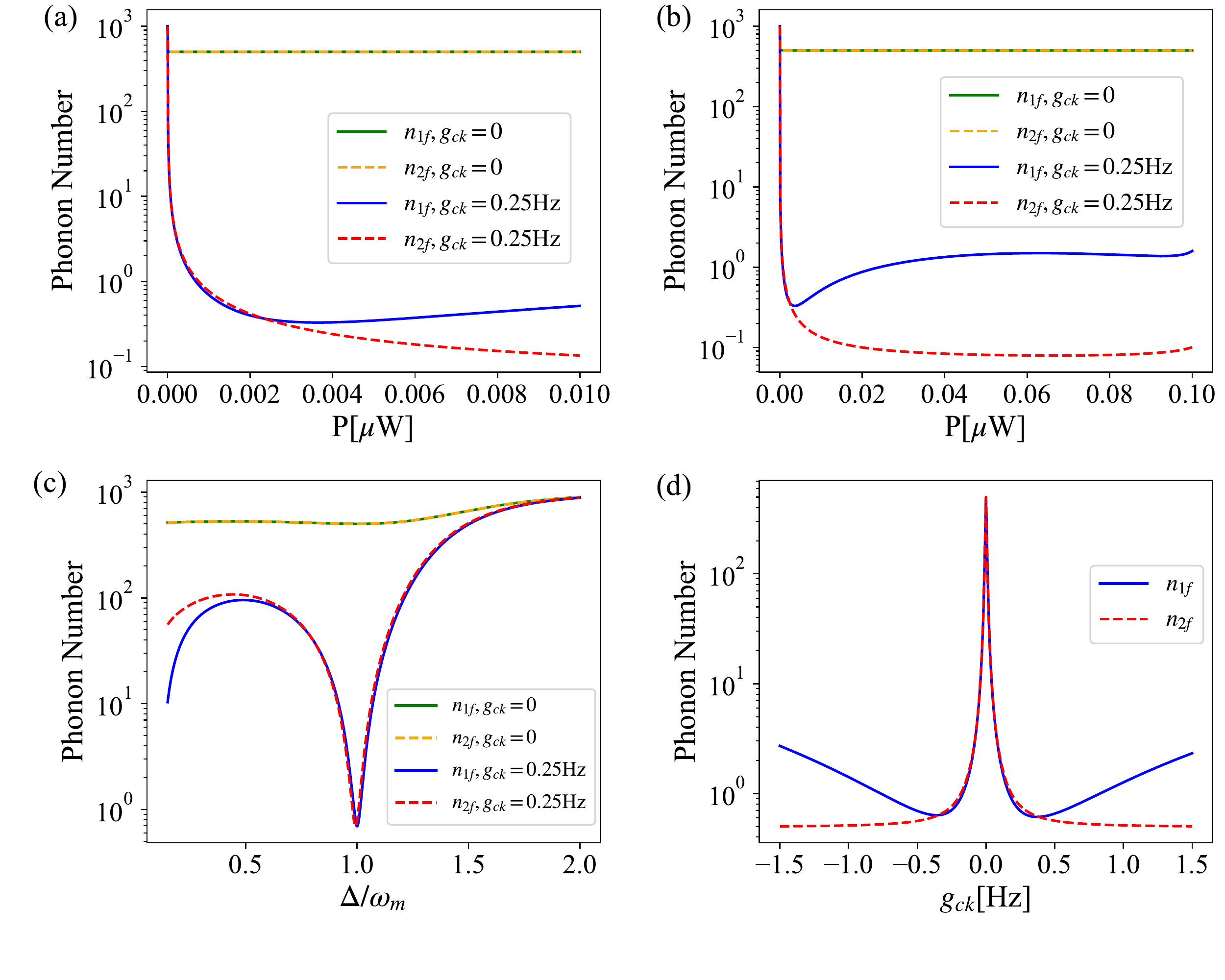}
    \caption{(a) Final average phonon number of mechanical mode 1 (2) versus input driving laser power with and without CK effect.$\Delta=\omega_{m}$. (b) Same to (a) with a larger scale of input driving laser power. $\Delta=\omega_{m}$. (c) Final average phonon number of mechanical mode 1 (2) versus detuning with and without CK effect. $P=10^{-3}\mu$W. (d) Final average phonon number of mechanical mode 1 (2) versus $g_{ck}$. $\Delta=\omega_{m}$,$P=10^{-3}\mu$W. Other parameters are same as those in Fig.\ref{multistates}. }
    \label{cooling}
    \end{figure}

    The cooling results are shown in Fig.\ref{cooling}. In Fig.\ref{cooling} (a) we investigate the effect of the input driving laser power on the cooling efficiency. One can find both mechanical mode 1 and 2 can not be cooled to ground state due to the existence of dark mode effect when there is no CK effect, i.e. the green and dashed orange curve in Fig.\ref{cooling} (a). However, when $g_{ck}=0.25$Hz, both mechanical mode 1 and 2 can be effectively cooled to ground state with the increase of input driving laser power. If we denote the critical laser power as the minimum power that makes the final phonon number $n_{f}\leqq 1$, one can find that the critical laser powers for mechanical mode 1 and 2 to reach ground state are 6.9$\times10^{-4}\mu$W and 7.6 $\times10^{-4}\mu$W, respectively. This slight difference originates from the CK effect between only mechanical mode 1 and the cavity mode. By increasing the region of the input laser power from 0.01 $\mu$W to 0.1 $\mu$W, we get the cooling result in Fig.\ref{cooling} (b). One can see that there is a minimum phonon number for mechanical mode 1, corresponding to the optimal input power 3.6$\times10^{-3}\mu$W. Similarly, there is also an optimal input power, i.e. 0.0665$\mu$W for mechanical mode 2 which is not so apparently shown in Fig.\ref{cooling} (b). Besides, one can see that mechanical 2 can always be cooled to ground state only if the input laser power exceeds 7.6 $\times10^{-4}\mu$W. While the mechanical mode 1 can no longer be cooled to ground state when the input laser power exceeds 0.0243 $\mu$W since $n_{1f}>1$ at this region. 
    
    The reason for this nonreciprocal cooling behavior is that there is only CK effect between mechanical 1 and the cavity mode. According to Eq.\ref{deltaprime}, $\Omega_{1}$ and $\Delta^{\prime}$ vary with the input power, so the difference $\vert \Omega_{1}-\Delta^{\prime} \vert $ and $\vert \omega_{2}-\Delta^{\prime} \vert $ vary, too. It's known that for one mechanical mode $\omega_{m}$ coupled to one optical mode $\omega_{c}$, the optimal cooling condition is the red detuning resonance, i.e. $\Delta=\omega_{c}-\omega_{L}=\omega_{m}$. Based on this, the cooling efficiency in our model varies with the input laser power. It's notable that the CK effect combined with the input laser breaks the dark mode effect here, since $\Omega_{1}$ deviates  from $\omega_{2}$ with an  amount of $g_{ck}\vert \alpha \vert^{2}$.  

    In Fig.\ref{cooling} (c) we demonstrate the steady final phonon number versus detuning $\Delta$. One can see that the ground state cooling fails in the absence of $g_{ck}$, which is shown in the green and dashed orange line in Fig.\ref{cooling} (c). When $g_{ck}=0.25$Hz, the ground state cooling of both mechanical mode 1 and 2 can be achieved in the vicinity of $\Delta/\omega_{m}=1$. However, the optimal $\Delta$ for cooling may not be $\Delta/\omega_{m}=1$ exactly any more. For example, the optimal detuning for the cooling of mechanical mode 2 is $\Delta/\omega_{m}=0.993$. This is caused by the modification of effective detuning $\Delta^{\prime}$ which has been illustrated in the last paragraph. In Fig.\ref{cooling} (d) we plot the steady final phonon number versus $g_{ck}$. One can see that both $n_{1f}$ and $n_{2f}$ are roughly symmetric to $g_{ck}=0$. The critical $g_{ck}$ for mechanical mode 1 (2) to reach ground state is $\pm 0.17$Hz  ($\pm 0.19$Hz), respectively. Still one can find a optimal $g_{ck}$ for mechanical mode 1 which has the same reason as that in Fig.\ref{cooling} (b).

    \begin{figure}
      \includegraphics[width=\linewidth]{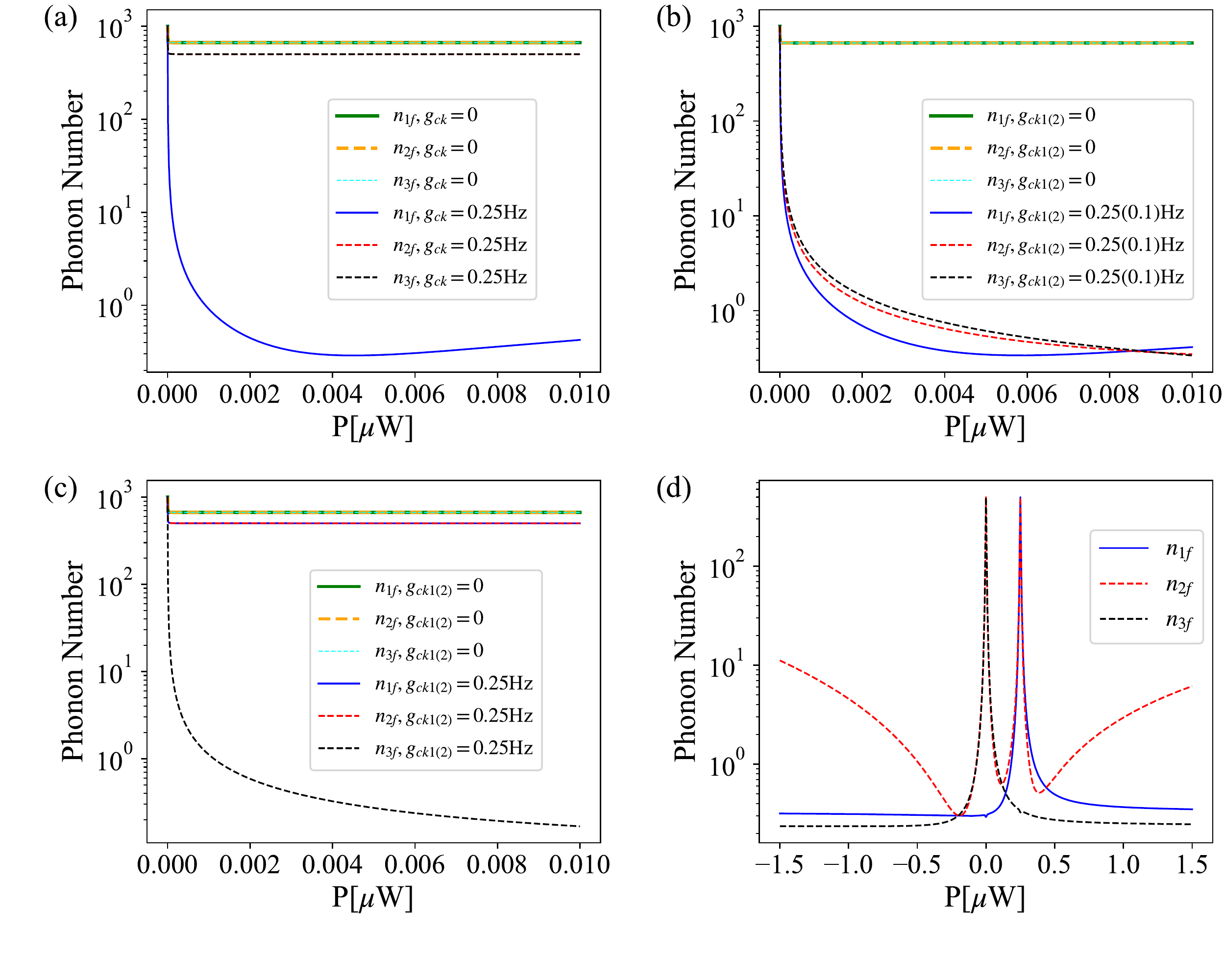}
      \caption{ The cooling effect of one cavity mode coupling to three degenerate mechanical modes. (a) The CK effect only exists between the cavity mode and mechanical mode 1. (b) Both mechanical mode 1 and mechanical mode 2 have CK effects. The CK coupling strengths are $g_{ck1}=0.25$Hz, $g_{ck2}=0.1$Hz, respectively. (c) is similar to (b) except that $g_{ck1}=g_{ck2}=0.25Hz$. (d) Phonon number versus $g_{ck2}$ with $g_{ck1}$ fixed at 0.25Hz. $P=4\times10^{-3}\mu$W. Other parameters are: $\Delta=\omega_{1}=\omega_{2}=\omega{3}=2\pi\times6.3$MHz,$\gamma_{1}=\gamma_{2}=\gamma_{3}=40$Hz,$g_{1}=g_{2}=g_{3}=250$Hz,$\kappa=2\pi\times0.1$MHz,$\omega_{L}=2\pi \times 1.31$GHz.  }
      \label{extension}
      \end{figure}

    We can also extend our model to the  optomechanical system with multiple degenerate mechanical modes. For example, we demonstrate the cooling effect of one optical mode coupling to three degenerate mechanical modes  in Fig.\ref{extension}. As is shown by the green,  dashed orange  and dashed cyan curve in Fig.\ref{extension} (a), one can find that all three mechanical modes can not be cooled to ground state due to the dark mode effect. When there is only CK effect between the mechanical mode 1 and the cavity mode, only the mechanical mode 1 can be cooled down to ground state, which is demonstrated by the blue,  dashed red and dashed black curve in Fig.\ref{extension} (a). The reason for the failure of the cooling of the mechanical mode 2 and 3 in Fig.\ref{extension}(a) is that they are still degenerate. To overcome this problem we can introduce the CK effect between the mechanical mode 2 and the cavity mode at the same time. The cooling result for $g_{ck1}$=0.25Hz, $g_{ck2}=0.1$Hz is shown in Fig.\ref{extension}  (b). One can see that all three mechanical modes are effectively cooled down to the ground state. However, it's notable that $g_{ck1}$ should not equal to $g_{ck2}$ otherwise there will be another dark mode effect between mechanical mode 1 and 2 which is shown in Fig.\ref{extension} (c). In Fig.\ref{extension} (d) we demonstrate the cooling effect for three mechanical modes with $g_{ck1}$ fixed at 0.25Hz while $g_{ck2}$ varying. One can see that there are two peaks at $g_{ck2}=0$Hz and $g_{ck2}=0.25$Hz. The left peak reveals the dark mode effect between the mechanical mode 2 and 3 while the right peak reveals the dark mode effect between the mechanical mode 1 and 2, respectively.

\section{CONCLUSION\label{con}}
   In this paper we propose a universal and scalable method to realize the simultaneous ground-state cooling of multiple degenerate mechanical modes by utilizing the CK effect. Due to the existence of the CK nonlinearity, there will be at most four stable steady states for optical and mechanical modes, which is more complex than the bistable behavior of the standard optomechanical system. In the following we take the result in branch 1 of the multiple stable steady states and investigate the cooling effect of the mechanical modes. For an optomechanical system consisting of two degenerate mechanical modes coupling to one optical mode, the simultaneous ground-state cooling for both mechanical modes can be effectively achieved by introducing the CK effect between only one mechanical mode and the cavity mode. A further study shows that there will be an optimal input laser power (the CK coupling strength $g_{ck}$) for cooling when the CK coupling strength $g_{ck}$ (input laser power) is fixed. This critical behavior is because that the effective detuning and mechanical resonant frequency are modulated by the solution of the steady-state equation  which is determined by $g_{ck}$ and input laser power. We can also extend our scheme to optomechanical system with more than two multiple degenerate mechanical modes coupling to one cavity mode. And we find that to realize the simultaneous cooling of N degenerate mechanical modes, the cavity mode should have CK couplings with N-1 mechanical modes whose strengths are different from each other. Our work can be considered as an effective scheme to manipulate the dark mode effect and quantum states in macroscopic systems.

  \begin{acknowledgments}
   
    We thank Dr.Guo-Qing Qin and Dr.Hao Zhang for helpful discussion. This work is supported by the  National Natural Science Foundation
    of China (61727801, 62131002), National Key Research and Development Program of China  (2017YFA0303700), the Key Research and Development  Program of Guangdong province (2018B030325002), Beijing Advanced Innovation Center for Future Chip (ICFC),
    and Tsinghua University Initiative Scientific Research  Program.
    
    \end{acknowledgments}

%

\end{document}